\documentclass[aoas,nameyear,seceqn,rotating,dvips]{arximspdf}
\usepackage{dcolumn}
\usepackage{graphicx}

\doi{10.1214/09-AOAS266}
\volume{3}
\issue{4}
\pubyear{2009}
\firstpage{1448}
\lastpage{1466}

\makeatletter
\newcolumntype{d}[1]{D{.}{.}{#1}}
\DeclareMathAlphabet\mathcaligr{OMS}{cmsy}{m}{n}
\makeatother

\begin{document}
\begin{frontmatter}

\title{Hierarchical mixture models for
assessing fingerprint individuality\thanksref{T1}}
\runtitle{Hierarchical Mixtures}

\begin{aug}
\author[A]{\fnms{Sarat C.} \snm{Dass}\ead[label=e1]{sdass@msu.edu}\ead[label=u1,url]{http://www.stt.msu.edu/\texttildelow sdass}\corref{}} \and
\author[B]{\fnms{Mingfei} \snm{Li}\ead[label=e2]{mli@bentley.edu}}
\runauthor{S. C. Dass and M. Li}
\affiliation{Michigan State University and Bentley University}
\address[A]{Department of Statistics \& Probability\\
Michigan State University\\
East Lansing,
Michigan 48824
\\ USA\\
\printead{e1}\\
\printead{u1}} 
\address[B]{Department of Mathematical Sciences\\
Bentley University\\
175 Forest Street\\
Waltham,
Massachusetts 025452\\
USA\\
\printead{e2}}
\thankstext{T1}{Supported by NSF Grant DMS-07-06385.}
\end{aug}

\received{\smonth{7} \syear{2008}}
\revised{\smonth{6} \syear{2009}}

%
\begin{abstract}
The~study of fingerprint individuality aims to determine to what extent
a fingerprint uniquely identifies an individual. Recent court cases
have highlighted the need for measures of fingerprint individuality
when a person is identified based on fingerprint evidence. The~main
challenge in studies of fingerprint individuality is to adequately
capture the variability of fingerprint features in a population. In
this paper hierarchical mixture models are introduced to infer the
extent of individualization. Hierarchical
mixtures utilize complementary aspects of mixtures at different
levels of the hierarchy. At the first (top) level, a mixture is used
to represent homogeneous groups of fingerprints in the population,
whereas at the second level, nested mixtures are used as
flexible representations of distributions of features from each
fingerprint. Inference for hierarchical mixtures is more challenging
since the number of unknown mixture components arise in both the
first and second levels of the hierarchy. A Bayesian
approach based on reversible jump Markov chain Monte Carlo
methodology is developed for the inference of all unknown parameters
of hierarchical mixtures. The~methodology is illustrated on
fingerprint images from the NIST database and is used to make inference
on fingerprint individuality estimates from this
population.
\end{abstract}

%

\end{frontmatter}
%

\section{Introduction}
\label{introduction}
Recent court cases have highlighted the need for
reporting error rates when an individual is identified based on
forensic evidence such as fingerprints. In the case of Daubert v.
Merrell Dow Pharmaceuticals [Daubert v. Merrell Dow Pharmaceuticals
Inc. (\citeyear{Da1993})], the U.S. Supreme Court ruled that in order
for expert
forensic testimony to be allowed in courts, it had to be subject to
five main criteria of scientific validation, that is, whether (i)
the particular technique or methodology has been subject to
statistical hypothesis testing, (ii) its error rates have been
established, (iii) standards controlling the technique's operation
exist and have been maintained, (iv) it has been peer reviewed, and
(v) it has a general widespread acceptance [see Pankanti, Prabhakar and
Jain (\citeyear{PaPrJa2002}) and Zhu, Dass and Jain (\citeyear
{ZhDaJa2007})]. Following Daubert, forensic
evidence based on fingerprints was first challenged in the 1999 case
of U.S. v. Byron C. Mitchell, stating that the fundamental premise
for asserting the uniqueness of fingerprints had not been objectively
tested and its potential matching error rates were unknown.
Subsequently, fingerprint based identification has been challenged in
more than 20 court cases in the United States. To address these
concerns, several research investigations have proposed measures that
characterize the extent of uniqueness of fingerprints (i.e.,
fingerprint individuality); see Pankanti, Prabhakar and Jain (\citeyear
{PaPrJa2002}),
Zhu, Dass and Jain (\citeyear{ZhDaJa2007}) and the references therein.
The~primary aim of these
measures is to capture the inherent variability and uncertainty when
an individual is identified based on fingerprint evidence.

The~statistical test of hypotheses for fingerprint based identification
can be set up as follows: Consider an input fingerprint with an
unknown identity $I_t$ being compared to the fingerprint of a
claimed identity $I_{c}$. The~test of hypotheses is
%
\begin{equation}
\label{hypothesistesting}
H_0\dvtx I_{t} \ne I_{c}\quad \mbox{versus} \quad H_1\dvtx
I_{t} =
I_{c},
\end{equation}
where $H_0$ (resp., $H_1$) is the hypothesis of a negative
(resp., positive) identification. The~hypotheses posed in the
order of negative vs. positive identification (as opposed to the
reverse order) allows us to control for the probability of making a
false positive identification (i.e., the probability of Type I
error). The~test of $H_0$ versus $H_1$ in (\ref{hypothesistesting})
is carried out by ascertaining the degree of similarity between the
two prints and involves two important steps: First, salient
fingerprint features are extracted from each print, and second, the
collection of features of the two prints are ``matched'' with each
other to obtain the best measure of similarity.

\begin{figure}[b]

\includegraphics{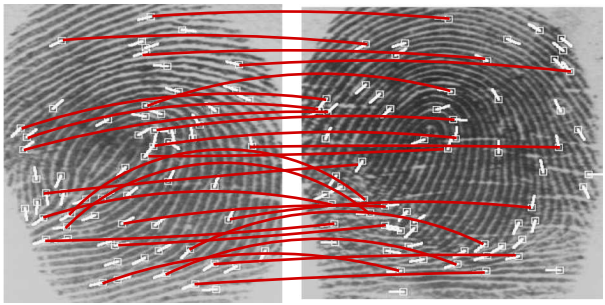}

\caption{Illustrating minutiae matching
[taken from Pankanti, Prabhakar
and Jain (\protect\citeyear{PaPrJa2002})]. A total of $m=64$ and $n=65$
minutiae were detected in left and right image, respectively, and
$25$ correspondences (i.e., matches) were found. The~white squares
and lines, respectively, represent the minutiae location and the
direction of ridge flow at that minutiae.}\label{fig:impostor:match}
\end{figure}

Figure \ref{fig:impostor:match} illustrates an example of the feature
extraction and matching procedures described in the previous
paragraph. Typical fingerprints as in Figure \ref{fig:impostor:match}
consist of smooth, nonintersecting flow patterns with alternating
dark and light lines, called ridges and valleys, respectively.
Occasionally, a ridge will either bifurcate or terminate and give
rise to an anomaly. The~anomalies in the ridge structures are called
minutiae which are the fingerprint features used for identifying
individuals. Figure \ref{fig:impostor:match} shows the locations of
minutiae ($x \in R^{2}$) as white squares for the two fingerprint
images extracted using a pattern recognition algorithm described in
Zhu, Dass and Jain (\citeyear{ZhDaJa2007}). Minutiae information of a
fingerprint is easy to
extract, permanent (does not change with time) and unique (distinct
minutiae patterns for different individuals), making it a popular
method for identifying individuals in the forensics community.
Subsequently, the number of matches is determined by an optimal rigid
transformation that brings the two sets of minutiae as close to each
other as possible and counting the number of minutiae in the right
panel that falls within a square of area $4r_0^{2}$ centered at each
minutiae in the left panel; $r_0$ is a small prespecified number
relative to the size of the fingerprint image. A higher number of
matches indicates a higher degree of similarity and favors the
rejection of $H_0$ in~(\ref{hypothesistesting}).

The~number of matching minutiae in Figure \ref{fig:impostor:match} is
$25$ but the question is: Should $H_0$ be rejected? In that case,
what is the uncertainty or error associated with the decision? This
is precisely the issue of fingerprint individuality since error rates
associated with the observed match are unknown. Pankanti, Prabhakar and
Jain (\citeyear{PaPrJa2002})
and Zhu, Dass and Jain (\citeyear{ZhDaJa2007}) propose using the
probability of a random
correspondence (PRC) as a measure of fingerprint individuality.
Mathematically, the PRC is expressed as
%
\begin{equation}
\label{PRC}
\mathrm{PRC}( w\vert m,n) = P(S \ge w\vert m,n ),
\end{equation}
where the random variable $S$ denotes the number of minutiae matches,
$w$ denotes the observed number of matches, and $m$ and $n$,
respectively, are the number of minutiae in the two fingerprint images.
The~probability in (\ref{PRC}) is calculated assuming $H_0$ is true,
that is, the pair of prints are impostors coming from two different
individuals. Small
(resp., large) values of the PRC indicate low (resp.,
high) levels of uncertainty which correspond to high (resp.,
low) extent
of fingerprint individualization. A PRC of $0.0004$, for example,
indicates that only $4$ out of $10{,}000$~impostor matches will result
in matching numbers that are greater than or equal to~$w$. So, having
observed $w$ causes us to suspect that $H_0$ may not be true.
The~uncertainty associated with this suspicion decreases as the
PRC gets smaller (i.e., closer to $0$). The~connection
between the PRC and the hypothesis testing criteria in
Daubert (which is one of the five main criteria for the scientific
validation of forensic evidence) can be seen as follows:
Under the hypotheses testing of~(\ref{hypothesistesting}),
the PRC is the $p$-value, computed under $H_0$, corresponding to
the observed number of matches $w$.

The~value of the PRC depends on the distribution of minutiae
locations in a pair of prints. Zhu, Dass and Jain (\citeyear{ZhDaJa2007})
demonstrated that when $m$ and $n$ are large, the distribution
of $S$ in (\ref{PRC}) can be
approximated by a Poisson distribution with mean (expected) number of
matches
%
\begin{equation}
\label{lambda}
\lambda(q_1,q_2,m,n) = m n p(q_1,q_2),
\end{equation}
where $q_{h}, h=1,2$ are the distributions fitted to the minutiae
locations in the pair of prints, and $p(q_1,q_2)$ is the probability of
a match given by
%
\begin{equation}
p(q_1,q_2) = \int\int_{(x,y)\dvtx x \in\mathcaligr{S}(y,r_0)}
q_{1}(x) q_{2}(y)\, dx\, dy,
\end{equation}
where $x\in R^{2}$ and $y\in R^{2}$ are independent minutiae from
$q_1$ and $q_2$, respectively, and $\mathcaligr{S}(y,r_0)$ is the square
of area $4r_0^{2}$ centered at $y$.

The~reliability of the PRC computed from a sample of fingerprints
depends on (1) how well elicited statistical models fit the
distribution of minutiae for different fingerprints, and (2) whether
the sample is representative of the target population. The~aim in
this paper is to develop methodology for (1) while implicitly
assuming the validity of (2). Thus, the results in Section
\ref{sec:fingerprint} are valid for a population which has the
fingerprint database as a representative sample.

It is well known, for example, that the distribution of minutiae
locations in fingerprints tend to form clusters [see, e.g.,
Scolve (\citeyear{Sc1979}), Stoney and Thornton (\citeyear{StTh1986})
and Zhu, Dass and Jain (\citeyear{ZhDaJa2007})].
Thus, candidate statistical models have to meet two important
requirements: (i) flexibility, that is, the model can represent a
variety of minutiae distributions for different fingerprints, and
(ii) associated measures of fingerprint individuality can be easily
obtained from these models. These considerations led Zhu, Dass and Jain
(\citeyear{ZhDaJa2007})
to propose mixture distributions as candidate choices for $q_1$ and
$q_2$. Based on mixtures of independent normals, the analytical
expression for $p(q_1,q_2)$ in (\ref{pmatch}) becomes
%
\begin{eqnarray}
p(q_{1},q_{2}) &=& 4 r_{0}^{2}
\sum_{k=1}^{K_{1}} \sum_{k'=1}^{K_{2}}
\prod_{b=1}^{2} \phi_{1}\bigl(0 |
\underbrace{\bigl(\mu_{k{1}}^{(b)}-\mu_{k{2}}^{(b)}\bigr)},
\underbrace{\bigl(\sigma_{k{1}}^{(b)}\bigr)^{2} +
\bigl(\sigma_{k{2}}^{(b)}\bigr)^{2}}\bigr),\\[-8pt]
\label{pmatch}&&
\hspace*{120pt}\mu\hspace*{65pt}
\sigma^{2}\nonumber
\end{eqnarray}
where $q_{h}(x) = \sum_{k=1}^{K_h}\prod_{b=1}^{2}
\phi_{1}(x^{(b)}\vert{\mu}_{kh}^{(b)},({\sigma
}_{kh}^{(b)})^{2})$
for $h=1,2$, $x=(x^{(1)},x^{(2)})$ and $\phi_{1}(\cdot\vert\mu
,\sigma^{2} )$
is the normal density with mean $\mu$ and variance $\sigma^2$.

One drawback of Zhu, Dass and Jain (\citeyear{ZhDaJa2007}) is that no
statistical model is
elicited on the minutiae for a \textit{population} of fingerprints;
standard mixture distributions were proposed for minutiae
distributions in each fingerprint separately. As a result, no
inference (e.g., confidence intervals) can be obtained for the
population version of the PRC. This is our motivation for developing
hierarchical mixture models and related inferential tools in this
paper. The~hierarchical mixture model [see
(\ref{hierarchicalmixture})] is a model on the minutiae for a
population of fingerprints that satisfies both requirements of (i)
flexibility and (ii) computational ease mentioned earlier. We assume
that the fingerprint population consists of $G$ homogeneous groups
with respect to the distribution of minutiae, with $q_{g}$ and
$w_{g}$, respectively, denoting the distribution of minutiae
locations and population proportion of the $g$th sub-population,
$g=1,2,\ldots,G$. For a fingerprint pair coming from the
sub-populations $g_1$ and $g_2$ with $1 \le g_1,g_2 \le G$, we have
$q_1 = q_{g_1}$ and $q_2 = q_{g_2}$ in (\ref{lambda}). Hence, it
follows that the population mean PRC corresponding to $w$ observed
matches in
the population is given by
%
\begin{equation}
\label{PRC2}
\overline{\mathrm{PRC}}(w\vert m,n ) =
\sum_{g_1=1}^{G} \sum_{g_2=1}^{G} \omega_{g_1} \omega_{g_2}
P\bigl(S\ge
w\vert\lambda(q_{g_1},q_{g_2},m,n) \bigr),
\end{equation}
where $S$ follows a Poisson distribution with mean
$\lambda(q_{g_1},q_{g_2},m,n)$.

In this paper a Bayesian framework for the inference from
hierarchical mixture models is developed, which in turn can be used
to make inference for the population mean PRC in (\ref{PRC2}).
Hierarchical mixture models contain an unknown number of mixture
components at two levels. Green (\citeyear{Gr1995}) and Green and Richardson
(\citeyear{GrRi1997}) developed the reversible jump Markov chain Monte Carlo
(RJMCMC) approach for estimating the unknown number of mixture
components by exploring the space of models of varying dimensions.
The~RJMCMC procedure developed in this paper generalizes the work of
Green and Richardson (\citeyear{GrRi1997}) to hierarchical mixture
models with two
levels of hierarchy. The~rest of the paper is organized as follows:
Section \ref{hierarchicalmixturemodels} develops hierarchical mixture
models for a heterogeneous population of objects (the objects are
fingerprints in our application). Sections \ref{bayesian} and~\ref
{sec:posteriorinference} develop the Bayesian and RJMCMC
framework for inference from hierarchical mixture models. Section
\ref{sec:fingerprint} discusses the application to fingerprint
analysis using PRCs.

\section{Hierarchical mixture models}
\label{hierarchicalmixturemodels} Consider an object, $\mathcaligr{O}$,
selected at random from a heterogenous population, $\mathcaligr{P}$,
with $G$ (unknown) groups. Let\vspace*{1pt} $X \equiv(x_1,x_2,x_3,\ldots)$
denote the observables on $\mathcaligr{O}$ where $x_{j} \equiv
(x_j^{(1)},x_{j}^{(2)},\ldots,x_{j}^{(d)})'$ is a $d$-variate random
vector in $R^{d}$. A hierarchical mixture model for the distribution
of $\mathcaligr{O}$ in the population is
%
\begin{equation}
\label{hierarchicalmixture} q(\underline{x}) =
\sum_{g=1}^{G} \omega_{g} \prod_{j=1}^{n} q_{g}(x_j),
\end{equation}
where $\underline{x} = (x_1,x_2,\ldots,x_n)$ are the $n$ observations
made on $\mathcaligr{O}$, $\omega_{g}, g=1,2,\ldots,G$ are the $G$
cluster proportions with $\omega_{g} > 0$ and
$\sum_{g=1}^{G}\omega_{g}=1$, $q_{g}(\cdot)$ is the mixture density
for the $g$th cluster given by
%
\begin{equation}
\label{standardmixture} q_{g}(x) =
\sum_{k=1}^{K_g} p_{kg} f_{kg}(x\vert\lambda_{kg}),
\end{equation}
with $f_{kg}$ denoting a density with respect to the Lebesgue measure
on $R^{d}$, $p_{kg}$ denoting the mixing probabilities satisfying:
(1) $p_{kg}
> 0$ and\vspace*{1pt} (2) $\sum_{k=1}^{K_g} p_{kg}=1$, and $\lambda_{kg}$
denoting the
set of all unknown parameters in $f_{kg}$. Identifiability of the
hierarchical mixture model of (\ref{hierarchicalmixture}) with
respect to its components is achieved by imposing the constraints
%
\begin{equation}
\label{thetaconstraints} \omega_{1} < \omega_{2} < \cdots<
\omega_{G}\quad \mbox{and}\quad \lambda_{1g} \prec\lambda_{2g}
\prec
\cdots\prec\lambda_{K_{g}g}
\end{equation}
for each $g=1,2,\ldots,G$, where $\prec$ is a partial ordering to be
defined later. The~set of all unknown parameters in the hierarchical
mixture model (\ref{hierarchicalmixture}) is denoted by $
\bolds{\theta} = (G,\bolds{\omega},\mathbf{K},\mathbf{p}
,\bolds{\lambda}), $ where $\bolds{\omega} =(\omega_1,\omega_2,\ldots
,\omega_{G})$, $\mathbf{K} =(K_1,K_2,\ldots,\break K_{g})$, $\mathbf{p} =
(p_{kg},k=1,2,\ldots,K_{g}, g=1,2,\ldots,G)$ and
$\bolds{\lambda} = (\lambda_{kg},k=1,2,\ldots,\break K_{g},
g=1,2,\ldots,G)$.

Hierarchical mixture models consists of two levels of hierarchy: At
the first (top, or $G$) level, the mixture is used to represent the
groups, whereas at the second (or $K_g$) level, nested mixture models
(nested within each $g=1,2,\ldots,G$ specification) are used as a
flexible representation of the distribution of observables. The~unknown
number of mixture components, or mixture complexity, arise at
both levels of the hierarchy, and is, therefore, more challenging to
estimate compared to standard mixtures. Estimating mixture complexity
has been the focus of intense research for many years, resulting in
various estimation methodologies in a broad application domain.
Nonparametric methods were developed in Escobar and West (\citeyear
{EsWe1995}) and
Roeder and Wasserman (\citeyear{RoWa1997}), whereas Ishwaran, James and
Sun (\citeyear{IsJaSu2001}) and Woo
and Sriram (\citeyear{WoSr2007}) developed methodology for the robust
estimation of
mixture complexity for count data. As discussed earlier, our approach
for estimating mixture complexity will be Bayesian based on the
RJMCMC algorithm.

In the subsequent text we assume each $f_{kg}$ is multivariate
normal with mean vector $\bolds{\mu}_{kg} \equiv
(\mu_{kg}^{(1)},\mu_{kg}^{(2)},\ldots,\mu_{kg}^{(d)})' \in R^{d}$ and
covariance matrix $\sum_{kg} \in R^{d}\times R^{d}$.\vspace*{1pt}

Our analysis on the fingerprint images in the NIST database (see
Section \ref{sec:fingerprint}) reveal that it is adequate to consider
diagonal covariance matrices of the form $\sum_{kg} =
\operatorname{diag}((\sigma_{kg}^{(1)})^{2},(\sigma
_{kg}^{(2)})^{2},\ldots,(\sigma_{kg}^{(d)}
)^{2})$,
where $(\sigma_{kg}^{(b)})^{2}$ is the variance of the\vspace*{2pt}
$b$th component. Four different choices of the covariance matrix
$\sum_{kg}$ are considered, namely, diagonal covariance matrix with
(i) common entries over $k$ [i.e., $\sigma_{kg}^{(d)} =
\sigma_{g}^{(d)}$, for some common value of $\sigma_{g}^{(d)}]$, (ii)
different entries over $k$, unrestricted covariance matrix with
(iii) common entries over $k$, and (iv) different entries over $k$.
These four choices are evaluated using the
Bayes Information Criteria (BIC) which is a model selection criteria
that favors parsimonious models consistent with the observed data.
The~highest BIC was found for the choice of diagonal covariance matrix of
(i) or (ii) for almost all of the fingerprints in the NIST database;
see Table~\ref{table:BIC}.

\begin{table}[b]
\vspace*{-2pt}\caption{Covariance matrix selection: Entries give
the number and percentages of fingerprint images in the NIST database
that ranked each covariance model as the top choice based on BIC}\label
{table:BIC}
\begin{tabular*}{\textwidth}{@{\extracolsep{\fill}}lccccc@{}}
\hline
\textbf{Covariance choice} & \textbf{(i)} & \textbf{(ii)} & \textbf
{(iii)} & \textbf{(iv)} & \textbf{Total}\\
\hline
Frequency & 1731 & 238 & 0 & 29 & 1998\\
Percentage & 86.64 & 11.91 & 0 & 1.45 & 100.00\\
\hline
\end{tabular*}
\end{table}

Thus, we take the density $f_{kg}$ in (\ref{standardmixture}) to be
%
\begin{equation}
f_{kg}(x\vert\lambda_{kg}) =
\phi_{d}(x\vert\bolds{\mu}_{kg},\bolds{\sigma}_{kg}) =
\prod_{b=1}^{d}
\phi_{1}\bigl(x^{(b)}\big\vert{\mu}_{kg}^{(b)},\bigl(\sigma
_{kg}^{(b)}\bigr)^{2}\bigr),
\end{equation}
where $\phi_{1}(\cdot\vert\mu,\sigma^{2})$ denotes the density of the
univariate normal distribution with mean $\mu$ and variance
$\sigma^{2}$, and $\bolds{\sigma}_{kg} \equiv
((\sigma_{kg}^{(1)})^{2},(\sigma
_{kg}^{(2)})^{2},\ldots,(\sigma_{kg}^{(d)}
)^{2})'$
is the $d$-variate\vspace*{1pt} vector of the variances. The~second
identifiability condition of (\ref{thetaconstraints}) is re-expressed
in terms of the first component of the mean vector as
%
\begin{equation}
\label{muconstraints0}
\mu_{1g}^{(1)} < \mu_{2g}^{(1)} < \cdots<
\mu_{K_{g}g}^{(1)}.
\end{equation}

For $N$ independent objects selected randomly from the population, it
follows that the distribution of observables for the $i$th object,
$i=1,2,\ldots,N$ has the density
%
\begin{equation}
\label{ithdensity}
q(\underline{x}_{i}) =
\sum_{g=1}^{G} \omega_{g} \prod_{j=1}^{n_i}
\sum_{k=1}^{K_g} p_{kg} \phi_{d}(x_{ij}\vert\bolds{\mu}_{kg},
\bolds{\sigma}_{kg}),
\end{equation}
where $\underline{x}_{i} \equiv( x_{ij}, j=1,2,\ldots,n_i )$ is
the set of $n_i$ observations made on the $i$th object with each
$x_{ij} \in R^{d}$, for $j=1,2,\ldots,n_i$. It follows from
independence that the joint distribution of all observables,
$\mathbf{x} \equiv(\underline{x}_{i}, i=1,2,\ldots,N )$, from
$N$ objects is given by $\prod_{i=1}^{N} q(\underline{x}_{i})$.

Two other notations are introduced here: $\bolds{\mu}$ and
$\bolds{\sigma}$ will respectively denote the collection of all
$\{\bolds{\mu}_{kg}, k=1,2,\ldots,K_g,g=1,2,\ldots,G\}$ and
$\{\bolds{\sigma}_{kg}, k=1,2,\break \ldots,K_g,g=1,2,\ldots,G\}$
vectors. Our goal is to infer the unknown parameters
$\bolds{\theta} =
(G,\bolds{\omega},\mathbf{K},\mathbf{p},\bolds{\mu},\bolds{\sigma})
$ based on the observed data $\mathbf{x}$.

\section{A Bayesian framework for inference}
\label{bayesian}
For the subsequent text, some additional notation is
introduced. The~symbol $I(\mathcaligr{S})$ will denote the indicator
function of the set $\mathcaligr{S}$, that is, $I(\mathcaligr{S})=1$ if
$\mathcaligr{S}$ is true, and $0$, otherwise. The~notation
$A,B,\ldots\vert C,D,\ldots$ will denote the distribution of random
variables $A,B,\ldots$ conditioned on $C,D,\ldots,$ with
$\pi(A,B,\ldots\vert C,D,\ldots)$ denoting the specific form of the
conditional distribution. Also, $\pi(A,B,\ldots\vert\cdot)$ will
denote the distribution of $A,B,\ldots$ given the rest of the
parameters. We specify a joint prior distribution on
$\bolds{\theta}$ in terms of the hierarchical specification
%
\begin{equation}
\label{priorx}
\pi(\bolds{\theta}) =
\pi(G, \mathbf{K})\cdot\pi(\bolds{\omega}, \mathbf{p}\vert G, \mathbf{K})
\cdot\pi(\bolds{\mu}\vert G, \mathbf{K})\cdot
\pi(\bolds{\sigma}\vert G, \mathbf{K}).
\end{equation}
The~component priors in (\ref{priorx}) are as follows:

\begin{longlist}
\item[(1)] The~prior on the mean vector is taken as
%
\begin{eqnarray}
\pi(\bolds{\mu}\vert\mathbf{K},G) &=&
\prod_{g=1}^{G} \Biggl[\Biggl(K_{g}!
\prod_{k=1}^{K_g} \phi_{1}\bigl(\mu_{kg}^{(1)}\vert\mu_0,\tau^{2}\bigr)
\Biggr)\nonumber\\
&&\hspace*{19pt}{}\times\bigl(I\bigl(\mu_{1g}^{(1)} < \mu_{2g}^{(1)} <
\cdots<
\mu_{K_{g}g}^{(1)}\bigr)\bigr)\\
\label{muprior}
&& \hspace*{37pt}{}\times\Biggl(\prod_{b=2}^{d}
\prod_{k=1}^{K_g}\phi_{1}\bigl(\mu_{kg}^{(b)}\vert\mu_0,\tau^{2}\bigr)
\Biggr)\Biggr].\nonumber
\end{eqnarray}
The~indicator function appears due to
the identifiability constraint (\ref{thetaconstraints}) imposed on
$\bolds{\mu}$ with resulting normalizing
constant $K_g!$ for each $g=1,2,\ldots,G$.

\item[(2)] The~prior distribution of the variances is taken as
%
\begin{equation}
\label{sigmaprior}
\pi(\bolds{\sigma}\vert\mathbf{K},G) =
\prod_{g=1}^{G} \Biggl(\prod_{k=1}^{K_g}
\prod_{b=1}^{d} \operatorname{IG}\bigl(\bigl(\sigma_{kg}^{(b)}\bigr
)^{2}\vert
\alpha_0,\beta_0\bigr)
\Biggr),
\end{equation}
where $\operatorname{IG}$ denotes the inverse gamma distribution with
prior shape and scale parameters $\alpha_0$ and $\beta_0$,
respectively.

\item[(3)] The~prior on the first and second level mixing proportions is
taken as
%
\begin{eqnarray}\label{prioromega}
\pi(\bolds{\omega}, \mathbf{p}\vert G, \mathbf{K}) &=&
G! D_{G}(\bolds{\omega}\vert\delta_{\omega})\cdot I(\omega
_1 <
\omega_2 < \cdots< \omega_{G})\nonumber\\[-8pt]\\[-8pt]
&&{}\times  \prod_{g=1}^{G}
D_{K_g}(\mathbf{p}_{g}\vert\delta_{p}),\nonumber
\end{eqnarray}
where
$D_{H}(\cdot\vert\delta)$ denotes the $H$-dimensional Dirichlet
density with the $H$-compo\-nent baseline measure
$(\delta,\delta,\ldots,\delta)$, where $\delta$ is a prespecified
constant, and
$\mathbf{p}_{g}\equiv(p_{1g},p_{2g},\ldots,p_{Kg,g})'$. The~indicator
function arises due to the imposed identifiability
constraint (\ref{thetaconstraints}) on $\bolds{\omega}$. It
follows that $G!$ is the appropriate normalizing constant for this
constrained density, obtained by integrating out
$\bolds{\omega}$ and noting that
$D_{G}(\bolds{\omega}\vert\delta_{\omega})$ is invariant under
different permutations of $\bolds{\omega}$.

\item[(4)] The~prior on $G$ and $\mathbf{K}$ is taken as
%
\begin{equation}
\label{priorG}
\pi(G,\mathbf{K})= \pi(G)\cdot
\pi(\mathbf{K}\vert G) = \pi_0(G)\cdot\prod_{g=1}^{G} \pi_0(K_g),
\end{equation}
where $\pi_0$ is the discrete uniform distribution between
$G_{\mathrm{min}}$ and $G_{\mathrm{max}}$ (resp., $K_{\mathrm{min}}$ to $K_{\mathrm{max}}$), both
inclusive, for $G$ (resp., $K_{g}$).
\end{longlist}

The~prior on $\bolds{\theta}$ depends on the hyper-parameters
$\delta_{p}$, $\delta_{\omega}$, $G_{\mathrm{max}}$, $G_{\mathrm{min}}$,
$K_{\mathrm{min}}$, $K_{\mathrm{max}}$, $\mu_0$, $\tau^{2}$, $\alpha_0$
and $\beta_0$,
all of which need to be specified for a given application. The~reader
is referred to our technical report [Dass and Li (\citeyear{DaLi2008})]
for these
specifications.

The~likelihood of the hierarchical mixture model involves several
summations within each product term and is simplified by augmenting
variables to denote the class labels of the individual observations.
Two different class labels are introduced for the two levels of
hierarchy: (1) The~augmented variable $\mathbf{W}\equiv
(W_1,W_2,\ldots,W_{N})$ denotes the class label of the $G$
sub-populations, that is, $W_{i} = g$ whenever object $i$ arises from
the $g$th subpopulation, and (2)~$\mathbf{Z} \equiv
({Z}_1,{Z}_{2},\ldots,{Z}_{N})$ with ${Z}_{i} \equiv
(Z_{ij},j=1,2,\ldots,n_i)$, where $Z_{ij} = k$ for $1\le k\le K_{g}$
if $x_{ij}$ arises from the $k$th mixture component
$\phi_{d}(\cdot\vert\bolds{\mu}_{kg},\bolds{\sigma}_{kg})$.
We denote the augmented parameter space by the same symbol
$\bolds{\theta}$ as before, that is, $ \bolds{\theta} =
(G,\bolds{\omega},\mathbf{K},\mathbf{p},\bolds{\mu},
\bolds{\sigma},\mathbf{W},\mathbf{Z})$. The~augmented
likelihood is now
\begin{eqnarray}\label{augmentedlikelihood}
&&\ell(G,\bolds{\omega},\mathbf{K},\mathbf{p},\bolds{\mu},\bolds{\sigma
},\mathbf{W},\mathbf{Z})\nonumber\\[-8pt]\\[-8pt]
&&\qquad=\prod_{i=1}^{N} \prod_{j=1}^{n_i} \prod_{g=1}^{G} \prod
_{k=1}^{K_g}
(\phi_{d}(x_{ij}\vert\bolds{\mu}_{kg},\bolds{\sigma
}_{kg}))^{I(Z_{ij}=k,W_{i}=g)},\nonumber
\end{eqnarray}
with priors on $\mathbf{W}$ and $\mathbf{Z}$ given by
%
\begin{equation}
\label{WandZdistribution}
\pi(\mathbf{W},\mathbf{Z}\vert G,\mathbf{K},
\bolds{\omega},\mathbf{p} )
= \pi(\mathbf{W}\vert G, \bolds{\omega})\cdot
\pi(\mathbf{Z}\vert G, \mathbf{K}, \mathbf{W},\mathbf{p} ),
\end{equation}
where $\pi(\mathbf{W}\vert G,\bolds{\omega}) =
\prod_{i=1}^{N}\prod_{g=1}^{G} \omega_{g}^{I(W_i = g)}$ and
\[
\pi(\mathbf{Z}\vert G, \mathbf{K}, \mathbf{W},\mathbf{p} )
= \prod_{g=1}^{G}\prod_{i\dvtx W_{i}=g}\prod_{j=1}^{n_i} \prod
_{k=1}^{K_g} p_{kg}^{I(Z_{ij}=k)}.
\]
Based on the augmented likelihood and prior distributions, one can
write down the posterior distribution (up to a normalizing constant)
via Bayes theorem. The~posterior has the expression
\begin{eqnarray}\label{posterior}
\pi(\bolds{\theta}\vert\mathbf{x} )
&\propto&
\ell(G,\bolds{\omega},\mathbf{K},\mathbf{p},\bolds{\mu},\bolds{\sigma
},\mathbf{W},\mathbf{Z})
\times
\pi(\mathbf{W},\mathbf{Z}\vert G,\mathbf{K},
\bolds{\omega},\mathbf{p})\nonumber\\[-8pt]\\[-8pt]
&&{}\times
\pi(G,\mathbf{K},\bolds{\omega},\mathbf{p},
\bolds{\mu},\bolds{\sigma})\nonumber
\end{eqnarray}
based on
(\ref{priorx}), (\ref{augmentedlikelihood}),
(\ref{WandZdistribution}) and observed data $\mathbf{x}$.

\section{Posterior inference} \label{sec:posteriorinference}
The~total number of unknown parameters in the hierarchical mixture
model depends on the values $G$ and $\mathbf{K}$. Thus, the
posterior in (\ref{posterior}) can be viewed as a probability
distribution on the space of all hierarchical mixture models with
varying dimensions. To obtain posterior inference for such a space of
models, Green (\citeyear{Gr1995}) and Green and Richardson (\citeyear
{GrRi1997}) developed the
RJMCMC for Bayesian inference. In this paper we develop a RJMCMC
approach to explore the posterior distribution in (\ref{posterior})
resulting from the hierarchical mixture model specification. We
briefly discuss the general RJMCMC implementation here. Let
$\bolds{\theta}$ and $\bolds{\theta}^{*}$ be elements of
the model space
with possibly differing dimensions. The~RJMCMC approach proposes a
move, say, $m$, with probability~$r_{m}$. The~move $m$ takes
$\bolds{\theta}$ to $\bolds{\theta}^{*}$ via the proposal
distribution
$q_{m}(\bolds{\theta},\bolds{\theta}^{*})$. In order to
maintain the time
reversibility condition, we require to accept the proposal with
probability
%
\begin{equation}
\label{acceptanceprobability}
\alpha(\bolds{\theta},\bolds{\theta}^{*}) =
\min\biggl\{1,
\frac{\pi(\bolds{\theta}^{*}\vert\mathbf{x})}{\pi
(\bolds{\theta}\vert\mathbf{x})}
\frac{r_{m'}q_{m'}(\bolds{\theta}^{*},\bolds{\theta
})}{r_{m}q_{m}(\bolds{\theta},\bolds{\theta}^{*})}
\biggr\};
\end{equation}
in (\ref{acceptanceprobability}),
$q_{m'}(\bolds{\theta}^{*},\bolds{\theta})$ represents the
probability of moving from $\bolds{\theta}^{*}$ to
$\bolds{\theta}$ based on the ``reverse'' move $m'$, and
$\pi(\bolds{\theta}\vert\mathbf{x})$ denotes the posterior
distribution of $\bolds{\theta}$ given $\mathbf{x}$. It is
crucial that the moves $m$ and $m'$ be reversible [see Green (\citeyear
{Gr1995})],
meaning that the densities
$q_{m}(\bolds{\theta},\bolds{\theta}^{*})$ and
$q_{m'}(\bolds{\theta}^{*},\bolds{\theta})$ have the same
support with respect to a dominating measure. In case
$\bolds{\theta}^{*}$ represents the higher dimensional model, we
can first sample $\mathbf{u}$ from a proposal
$q_0(\bolds{\theta},\mathbf{u})$ (with possible dependence
on $\bolds{\theta}$), and then obtain $\bolds{\theta}^{*}$
as a one-to-one function of $(\bolds{\theta},\mathbf{u})$.
In that case, the proposal density
$q_{m}(\bolds{\theta},\bolds{\theta}^{*})$ in
(\ref{acceptanceprobability}) is expressed as
%
\begin{equation}
\label{q}
q_{m}(\bolds{\theta},\bolds{\theta}^{*}) =
q_0(\bolds{\theta},\mathbf{u})\big/\operatorname{det}\biggl[\frac
{\partial
\bolds{\theta}^{*}}{\partial(\bolds{\theta},\mathbf{u})}\biggr],
\end{equation}
where $\frac{\partial\bolds{\theta}^{*}}{\partial
(\bolds{\theta},\mathbf{u})}$ denotes the Jacobian of the
transformation from $(\bolds{\theta},\mathbf{u})$ to
$\bolds{\theta}^{*}$, and $\operatorname{det}$ represents the absolute
value of
its determinant. If the triplet $(\bolds{\theta},\mathbf{u},\bolds{\theta}^{*})$
involves some discrete components, then the Jacobian of the
transformation is obtained by the one-to-one map of the continuous
parts of $\bolds{\theta}^{*}$ and $(\bolds{\theta
},\mathbf{u})$,
which can depend on the values realized by the discrete components.

For the inference on hierarchical mixture models, five types of
updating steps are considered with reversible pairs of moves,
$(m,m')$, corresponding to moves in spaces of varying dimensions.
The~outline of the steps are as follows:
%
\begin{equation}
\label{updatingsteps}
\cases{
\mbox{(1) Update $G$ with $(m,m')\equiv(\mbox{$G$-split},\mbox
{$G$-merge})$},\cr
\mbox{(2) Update $\mathbf{K}\vert G, \bolds{\omega},
\mathbf{W}$ with $(m,m')\equiv
(\mbox{$K$-split},\mbox{$K$-merge})$},\cr
\mbox{(3) Update
$\bolds{\omega}\vert G,\mathbf{K},\mathbf{W},\mathbf{Z},
\mathbf{p},\bolds{{\mu}},\bolds{{\sigma}}$},\cr
\mbox{(4) Update
$\mathbf{W},\mathbf{Z}\vert G,\mathbf{K},\bolds{\omega},
\mathbf{p},\bolds{{\mu}},\bolds{{\sigma}}$ and}\cr
\mbox{(5) Update
$\mathbf{p},\bolds{{\mu}},\bolds{{\sigma}}\vert
G,\mathbf{K},\bolds{\omega},\mathbf{W},\mathbf{Z}$.}}
\end{equation}

Our methodological contribution is the development of the Update $G$
steps (\mbox{$G$-split} and $G$-merge) based on a pair of reversible jump
moves. The~steps for merging and splitting $G$ are described in
detail in the \hyperref[append]{Appendix}. The~Update $\mathbf{K}$
steps are similar
to that of Green and Richardson (\citeyear{GrRi1997}). The~other steps
\mbox{(3)--(5)} do not
involve jumps in spaces of varying dimensions, and can be carried out
based on a regular Gibbs proposal. One cycle through steps (1)--(5)
completes one iteration of the RJMCMC sampler.

The~assessment of convergence of the RJMCMC is carried out based on
the methodology of Brooks and Guidici (\citeyear{BrGi1998}, \citeyear
{BrGi2000}). A total of 3
chains are run from different starting points and different variance
components of the log-likelihood are calculated to obtain 3
diagnostic plots, namely, the plots of (i) the overall and within
chain variance, $\hat{V}$ and $W_{c}$, (ii) within model and within
chain within model variances, $W_{m}$ and $W_{m}W_{c}$, and (iii)
between model and between model within chain variances, $B_m$ and
$B_{m}W_{c}$, against the number of iterations. The~merging of the
two lines in each plot indicate that the chains have sufficiently
mixed.


\section{Assessing fingerprint individuality}
\label{sec:fingerprint}

\begin{figure}

\includegraphics{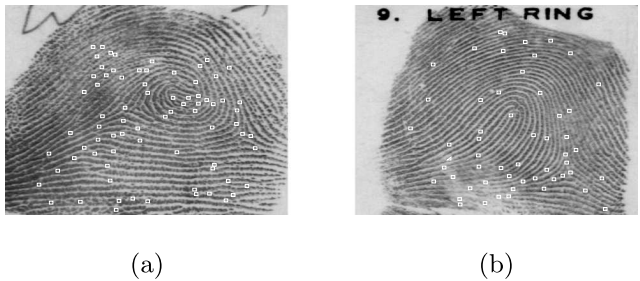}

\caption{Two fingerprint images from the NIST
database with minutiae locations indicated by white squares.}\label{fig:NIST}
\end{figure}

Our inferential methodology for assessing fingerprint individuality is
illustrated using fingerprint images from the \citeauthor{NIST}.
The~NIST fingerprint database is publicly available and
consists of 2000 8-bit gray scale fingerprint image pairs of size
512-by-512 pixels. Because of the similarity of the image pairs, only
the first image of each pair was used in the statistical modeling.
The~algorithm described in Zhu, Dass and Jain (\citeyear{ZhDaJa2007})
(also mentioned in the
\hyperref[introduction]{Introduction}) was used to extract minutiae
from these images;
minutiae could not be automatically extracted from two images of the
NIST database due to poor quality and these were discarded from
further consideration. Figure \ref{fig:NIST} shows examples of two
fingerprint images from the NIST database with minutiae locations
indicated by white squares.

\begin{figure}[b]

\includegraphics{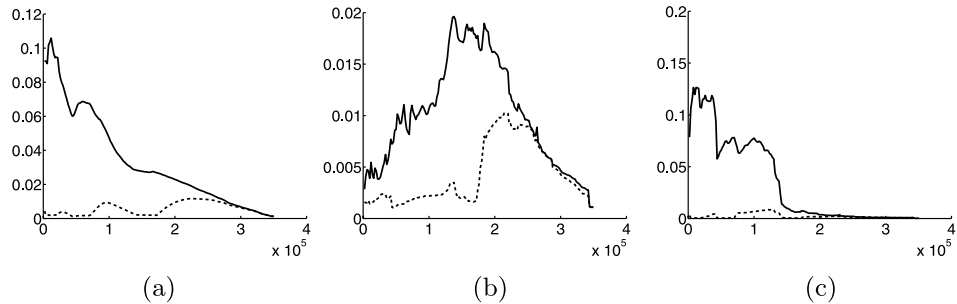}

\caption{Convergence diagnostics for the
NIST fingerprint database with $N_0=100$. Panels \textup{(a)}, \textup
{(b)} and \textup{(c)}, respectively,
show the plots of ($\hat{V},W_{c}$), ($W_m,W_mW_{c}$) and
($B_{m},B_mW_c$) as a function of the iterations. The~x-axis unit is
10,000 iterations.}\label{fig:NIST:convergence}
\end{figure}

\begin{figure}

\includegraphics{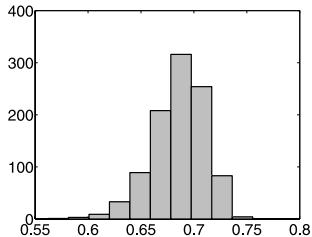}

\caption{\label{fig:PRC:hist} Posterior distribution of
$\overline{\mathrm{PRC}}$ based on $1000$ realizations of the RJMCMC after
$300{,}000$ iterations for $N_0=100$.}
\end{figure}

The~RJMCMC algorithm developed in the previous section is
used to obtain the posterior distribution of $\overline{\mathrm{PRC}}$.
The~first $N_0=100$ fingerprint images from the NIST database are taken as
the sample and three chains with starting values obtained using the
clustering procedure of Zhu, Dass and Jain (\citeyear{ZhDaJa2007}) are
run. Figure \ref
{fig:NIST:convergence} gives the
diagnostic plots of the RJMCMC sampler which establish convergence
after a burn-in of $B=250{,}000$ iterations. The~posterior distribution of
$\overline{\mathrm{PRC}}$ (corresponding to $m=64$, $n=65$, $w=25$ and
$r_0 =
15$ pixels) based on $1000$ realizations of the RJMCMC after the
burn-in period is given in Figure \ref{fig:PRC:hist} with a posterior
mean of $0.6859$ and the 95\% HPD interval given by
$[0.63,0.735]$. We conclude that if a fingerprint pair
was chosen from this population with $m=64$, $n=65$ and an observed
number of matches $w=25$, there is high uncertainty in making a
positive identification. Our analysis actually indicates that the
fingerprints in Figure~\ref{fig:impostor:match} represent a typical
impostor pair. The~95\% HPD set suggests that the PRC can be as high
as $0.735$, that is, about $3$ in every $4$ impostor pairs result in
$25$ or more matches.

How many matches does it take to positively identify an individual?
Different countries around the world have different standards [Girard
(\citeyear{Gi2007})]. In the Netherlands, this number is $12,$ whereas
in South
Africa, it is $7$. In the United States and the UK, this number is
not fixed and depends on expert testimonial. To assess the level of
uncertainty associated with these standards, we conduct a study of
the PRC based on $w=7$ matches. The~best case scenario corresponds to
the combination $(m,n,w) = (7,7,7)$ (when all query and template
minutiae match with each other) with a mean PRC of $5.09 \times
10^{-5}$ in Table \ref{table:7matches}. Note that 7 matches has
moderate strength of evidence for declaring a positive match; the PRC
implies 5 in 100,000 impostor fingerprint pairs will have all 7
minutiae match with each other. It is also very unlikely that $n=7$
in real life since fingerprints lifted from a crime scene have far
lesser number of minutiae (thus, $m \ll n$) compared to the template
it is being matched to. In this latter case, the PRCs are far larger
(see Table \ref{table:7matches}), making the case for positive
identification even weaker.

\begin{table}[b]
\caption{Mean PRCs for the combinations $(7,7,n)$}\label{table:7matches}
\begin{tabular*}{\textwidth}{@{\extracolsep{\fill}}lcccccc@{}}
\hline
$n$ & $7$ & $10$ & $15$ & $55$ & $65$ & $75$\\
Mean PRC& $5.09 \times10^{-5}$ & $1.40 \times10^{-4}$ & $3.25 \times
10^{-4}$ & $0.0155$ & $0.0333$ & $0.0614$ \\
\hline
\end{tabular*}
\end{table}

\begin{figure}

\includegraphics{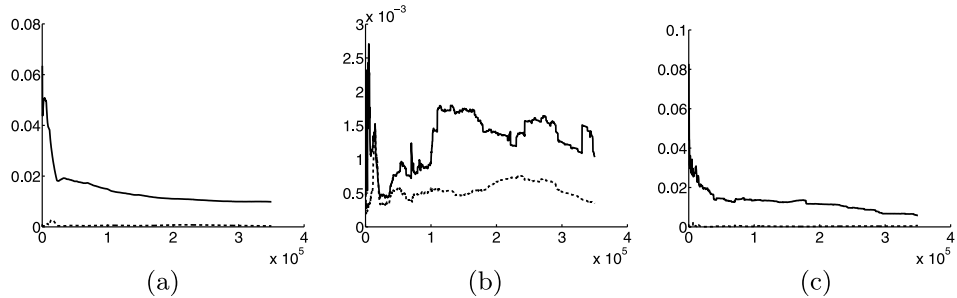}

\caption{Convergence diagnostics for the
NIST fingerprint database with $N_0=200$. Panels \textup{(a)}, \textup
{(b)} and \textup{(c)}, respectively,
show the plots of ($\hat{V},W_{c}$), ($W_m,W_mW_{c}$) and
($B_{m},B_mW_c$) as a function of the iterations. The~x-axis unit is
10,000 iterations.}\label{fig:NIST:convergence:200}
\end{figure}

To compare the results of inference using a larger sample size, we
ran the \mbox{RJMCMC} sampler for the first $N_0=200$ and $500$ fingerprint
images in the NIST database. The~computational complexity increases
in two ways: first, it takes longer, on the average, to complete one
iteration of the RJMCMC and second, the RJMCMC takes a longer time
to converge. On our personal computer with processing speed $2.66$
GHz and $1.96$ GB of RAM, it took about $12.5$, $31.8$ and~$90.0$
hours, respectively, to generate every $50{,}000$ iterations of the
RJMCMC for $N_0=100, 200$ and~$500$. While the RJMCMC converged at
$300{,}000$ iterations for $N_0=100$, the chain did not converge even
at $B=350{,}000$ iterations for $N_0=200$ (see Figure
\ref{fig:NIST:convergence:200}) and $N_0=500$ (the diagnostic plots
are not shown).

\textit{Discussion}: The~RJMCMC sampler is able to accommodate all
$N_0=1998$ fingerprint images from the NIST database. However, this
chain is extremely slow at mixing, and therefore, we do not expect
convergence to occur in real time on our computers. Computational
demand magnifies exponentially for very large databases such as the
US-VISIT program. Thus, for implementation on very large databases,
results can be obtained with the help of high-end computing
facilities.

One enormous advantage of the methodology outlined in this paper is
that the RJMCMC needs to be run on large databases \textit{only once}.
After convergence is achieved, inference on $\overline{\mathrm{PRC}}$
for any
combination of $(w,m,n)$ can be obtained using formula (\ref{PRC2}).
As an illustration based on our smaller sample size of $N_0=100$,
Table \ref{table:PRC:all} gives the results of this analysis for
different combinations of $(w,m,n)$. The~entries of Table
\ref{table:PRC:all} provides a general guideline to FBI and forensic
experts on the extent of uncertainty associated with making a
positive identification. Note that when $(w,m,n)=(25,64,65)$, the PRC
is high, indicating a low extent of individualization. However, Table
\ref{table:PRC:all} also provides several combinations of $(w,m,n)$
that favor positive identification with a high degree of
individualization. For example, we can look at entries in Table
\ref{table:PRC:all} for 95\% HPD sets that fall entirely below a
threshold, say, $T_0$. With the choice of $T_0=0.003$ (that is, 3 in
every $1000$ impostor fingerprint pairs will have $w$ or more
observed matches), the combinations that allow for positive
identification with uncertainty level of at most $T_0$ are
$(45,54,55)$, $(50,54,55)$, $(53,54,55)$, $(50,64,65)$ and
$(53,64,65)$; for these combinations, the probability that the true
PRCs occur below $T_0$ is at least 95\%. For larger values of~$N_0$,
the size of the HPD sets will decrease due to decreasing variability
of the estimate of $\overline{\mathrm{PRC}}$.

\begin{table}
\caption{Posterior means and $95\%$ HPD sets
calculated based on $1000$ realizations of~the~RJMCMC~for~$N_0=100$}\label{table:PRC:all}
\begin{tabular*}{10cm}{@{\extracolsep{\fill}}lcd{2.17}@{}}
\hline
$\bolds{w}$ & \multicolumn{1}{c}{\textbf{Mean}} & \multicolumn{1}{c}{\textbf{HPD}}\\
\hline
& \multicolumn{2}{c@{}}{$(m,n)=(54,55)$} \\
$25$ & $1.33 \times10^{-1}$ & (1.16,1.53) \times10^{-1}  \\
$35$ & $1.70\times10^{-3}$ & (0.80,3.40)\times10^{-3}\\
$45$ & $5.69\times10^{-4}$ & (0.0007,2.40) \times10^{-3}\\
$50$ & $5.68 \times10^{-4}$ & (0.00002,2.40) \times10^{-3}  \\
$53$ & $5.68\times10^{-4}$ & (0.0002,7.34) \times10^{-4} \\[5pt]
& \multicolumn{2}{c@{}}{$(m,n)=(64,65)$}\\
$25$ &  $6.84\times10^{-1}$ & (6.27,7.26) \times10^{-1} \\
$35$ & $9.22\times10^{-2}$ & (0.77,1.09)\times10^{-1}\\
$45$ &  $1.90\times10^{-3}$ & (0.93,3.60) \times10^{-3}  \\
$50$ & $6.42\times10^{-4}$ & (0.0051,2.40)\times10^{-3}  \\
$53$ & $5.80\times10^{-4}$ & (0.0089,2.40) \times10^{-3}  \\[5pt]
&  \multicolumn{2}{c@{}}{$(m,n)=(74,75)$} \\
$25$ &  $9.50 \times10^{-1}$&(8.99,9.87)\times10^{-1}\\
$35$ &  $6.10 \times10^{-1}$&(5.54,6.57) \times10^{-1}\\
$45$ &  $9.91\times10^{-2}$&(0.81,1.20)\times10^{-1} \\
$50$ &  $2.12 \times10^{-2}$ &(1.63,2.73)\times10^{-2} \\
$53$ &  $7.10 \times10^{-3}$ &(5.10,9.80) \times10^{-3} \\
\hline
\end{tabular*}
\end{table}

In this paper we only considered a two level hierarchical mixture model.
The~US-VISIT program now requires individuals to submit prints from all
10 fingers. This is the case of a 3-level hierarchical mixture model;
in the first (top) level, individuals form the $G$ groups based on
similar characteristics of their 10 fingers, and the distribution of
features in each finger is modeled using standard mixtures. Any higher
level hierarchical mixture models will be more involved in two ways: (1)
The~computational costs, including memory and time, since convergence
will be much slower to achieve, and (2) the development of reversible moves
such as $G$-merge and $G$-split for the higher level of mixtures. We
are of
the view that the best estimate of the population PRC can be obtained
if the
data is characterized by a model that best represents the way the data is
structured and observed. In the case of the US-VISIT, the 3-level hierarchical
mixture model is indeed the right way to view the available data. Further
research will be needed to see how the computational complexity can be reduced.
The~availability of high-end computing facilities will definitely be a
requirement
for fitting higher level hierarchical mixtures.

The~central issue for extending the proposed analysis to other
biometrics, such as face and iris, is the type of feature extracted
for each of the different biometrics. The~framework of hierarchical
mixture models will apply to these biometric traits but we have to
develop mixture models on different feature spaces. The~features we
used in this paper were minutiae locations, and therefore, we needed
mixture models on points in $R^{2}$. An additional feature for
fingerprints are the minutiae directions (the white lines in Figure
\ref{fig:impostor:match}). In order to run a similar analysis, one
would need to develop suitable mixture models on the product space
$R^{2}\times[0,2\pi)$. Similarly, in the case of iris, the feature
used is the IrisCode (consisting of a rectangular array of $0$s and
$1$s), and so the statistical models that have to be developed are
potentially Markov Random Field models (since there is significant
spatial dependence between neighboring $0$s and $1$s) indexed by a
set of parameters. Then, one could postulate that the population
consists of $G$ such groups of MRF models. We will also need a
distribution for the number of matching features and derive the
distribution of this under impostor pairs of IrisCodes.


\section{Summary and future work}
We have developed Bayesian inference methodology for hierarchical
mixture models with application to fingerprint individuality. One way
to further reduce the level of uncertainty for a fixed combination
$(w,m,n)$ is to increase the number of features used for matching.
Our future work will be to derive hierarchical mixture models on the
extended feature space consisting of minutiae locations and
directions. The~challenge here is that the angles are significantly
spatially correlated and the minutiae locations exhibit clusters. We
are currently developing a model that can account for these minutiae
characteristics. We plan to improve our algorithm so that it can be
run more quickly on very large databases. Hierarchical mixture models
have potential use in other areas as well, including the clustering
of soil samples (objects) based on soil characteristics which can be
modeled by a mixture or a transformation of mixtures.

\begin{appendix}\label{append}
\section*{Appendix}

In the subsequent text, the identifiability condition
(\ref{muconstraints0}) based on the first components of
$\bolds{\mu}_{kg}$ for $k=1,2,\ldots,K_g$ will be rewritten
using the `$\prec$' symbol as
%
\begin{equation}
\label{muconstraintvector}
\bolds{\mu}_{1g} \prec
\bolds{\mu}_{2g} \prec\cdots\prec\bolds{\mu}_{K_{g}g}
\end{equation}
for each $g=1,2,\ldots,G$. Let $\bolds{\theta}$
and $\bolds{\theta}^{*}$ denote two different states of the model
space, that is,
\begin{eqnarray}
\label{xyspace}
\bolds{\theta} &=&
(G,\bolds{\omega},\mathbf{K},\mathbf{p},\bolds{\mu},
\bolds{\sigma},\mathbf{W},\mathbf{Z})
\quad\mbox{and}\nonumber\\[-8pt]\\[-8pt]
\bolds{\theta}^{*} &=&
(G^{*},\bolds{\omega}^{*},\mathbf{K}^{*},\mathbf{p}^{*},\bolds{\mu}^{*},
\bolds{\sigma}^{*},\mathbf{W}^{*},\mathbf{Z}^{*}),\nonumber
\end{eqnarray}
where the ${}^*$s in (\ref{xyspace}) denote a possibly different
setting of the parameters.

\subsection{The $G$-merge move}
The $G$-merge move changes the current $G$ to $G-1$ (that is, $G^{*}
= G-1$) and is carried out based on the following steps:
{\smallskipamount=0pt
\begin{longlist}
\item[Step 1:] Two of the $G$ components, say, $g_1$ and $g_2$, with $g_1 <
g_2$, are selected randomly for merging into $g^{*}$ with $\omega
_{g^{*}} = \omega_{g_1} + \omega_{g_2}$.

\item[Step 2:] The $K$-components, $K_{g_1}$ and $K_{g_2}$, are combined to
obtain $K_{g^*}$ in the following way. Adding
$K_{g_1}+K_{g_2}=K_{t}$, we set
$K_{g^*} = (K_t+1)/2$ if $K_{t}$ is odd, and $K_{g^{*}} = K_t/2$ if
$K_t$ is even.

\item[Step 3:] Next, $(\mathbf{p}_{g_1},\bolds{\mu}_{g_1},\bolds{\sigma
}_{g_1})$ and $(\mathbf{p}_{g_2},\bolds{\mu}_{g_2},\bolds{\sigma
}_{g_2})$ are merged to obtain $(\mathbf{p}_{g^*},\break \bolds{\mu
}_{g^*},\bolds{\sigma}_{g^*})$
as follows. The identifiability conditions of (\ref
{muconstraintvector}) hold for $g=g_1$ and $g=g_2$, and must be ensured
to hold for $g=g^{*}$ after the merge step. To achieve this, the
$K_{t}$ $\bolds{\mu}$'s are arranged in increasing order
%
\begin{equation}
\label{mumergeincreasing}
\bolds{\mu}_{1} \prec\bolds{\mu}_{2} \prec\cdots\prec\bolds{\mu
}_{K_t-1} \prec\bolds{\mu}_{K_t}
\end{equation}
with associated probability $p_{j}$ for $\bolds{\mu}_{j}$, for
$j=1,2,\ldots,K_{t}$. Thus, $p_{j}$ are a re-arrangement of the $K_t$
probabilities in $\mathbf{p}_{g_1}$ and $\mathbf{p}_{g_2}$
according to the partial ordering on $\bolds{\mu}_{g_1}$ and
$\bolds{\mu}_{g_2}$ in (\ref{mumergeincreasing}). First, the
case when $K_{t}$ is even is considered. Adjacent $\bolds{\mu}$\vspace*{1pt}
values in (\ref{mumergeincreasing}) are paired
%
\begin{equation}
\label{mumergeincreasingstep}
\underbrace{\bolds{\mu}_{1} \prec\bolds{\mu}_{2}}
\prec\underbrace{\bolds{\mu}_{3} \prec\bolds{\mu}_{4}}
\prec\cdots\prec\underbrace{\bolds{\mu}_{K_{t}-1} \prec\bolds{\mu}_{K_t}}
\end{equation}
and the corresponding $g^{*}$ parameters
are obtained using the formulas $p_{kg^{*}}^{*} = \frac{p_{2k-1} +
p_{2k}}{2}$,
\begin{eqnarray}
\label{mergedparameters}
\bolds{\mu}_{kg^{*}}^{*}& =&
\frac{p_{2k-1}\bolds{\mu}_{2k-1} +
p_{2k}\bolds{\mu}_{2k}}{p_{2k-1}+p_{2k}} \quad\mbox{and}\nonumber\\[-8pt]\\[-8pt]
\bolds{\sigma}_{kg^{*}}^{*} &=&
\frac{p_{2k-1}\bolds{\sigma}_{2k-1} +
p_{2k}\bolds{\sigma}_{2k}}{p_{2k-1}+p_{2k}}\nonumber
\end{eqnarray}
for $k=1,2,\ldots,K_{g^{*}}$. To obtain $\mathbf{W}^{*}$ and
$\mathbf{Z}^{*}$, objects with $W_{i} = g_1$ or $W_i = g_2$ are
relabeled as $W_{i}^{*} = g^{*}$. For these objects,
the allocation to the $K_{g^*}$ components is carried out
using a Bayes allocation scheme. Explicit expressions for the
allocation probabilities are provided in Dass and Li (\citeyear{DaLi2008}).
When $K_{t}$ is odd, an index, $i_{0}$ is selected at
random from the set of all odd integers up to $K_{t}$, namely,
$\{1,3,5,\ldots, K_t\}$. The triplet
(${p}_{i_{0}},\bolds{\mu}_{i_{0}},\bolds{\sigma}_{i_{0}}$)
is not merged with any other indices but the new $p_{i_{0}}^{*} =
p_{i_{0}}/2$. The remaining adjacent indices are merged according to {Step~3}.
\end{longlist}}

\subsection{The $G$-split move}
The split move is reverse to the merge step above and is carried out
in the following steps:
{\smallskipamount=0pt
\begin{longlist}
\item[Step 1:] A candidate $G$-component for split, say, $g$, is
chosen randomly with probability $1/G$. The split components are
denoted by $g_1$ and $g_2$. The first level mixing probability,
$\omega_g$, is split into $\omega_{g_1}$ and $\omega_{g_2}$ by
generating a uniform random variable, $u_{0}$, in $[0,1]$ and setting
$ \omega_{g_1} = u_0
 \omega_{g}\mbox{ and }\omega_{g_2} = (1-u_0)  \omega_{g}.$

\item[Step 2:] The value of $K_{g}$ is transformed to $K_{t}$
where $K_{t}$ is either $2K_g-1$ or $2K_g$ with probability $1/2$
each. Once $K_{t}$ is determined, a pair of indices
$(K_{g_1},K_{g_2})$ is selected randomly from the set of all possible
pairs of integers in $\{ K_{\mathrm{min}},K_{\mathrm{min}}+1,\ldots,K_{\mathrm{max}}\}^{2}$
satisfying $K_{g_1}+K_{g_2}=K_{t}$. If $M_0$ is the total number of
such pairs, then the probability of selecting one such pair is
$1/M_0$. The selection of $K_{g_1}$ and $K_{g_2}$ determines the
number of second level components in the $g_1$ and $g_2$ groups.

\begin{figure}[b]

\includegraphics{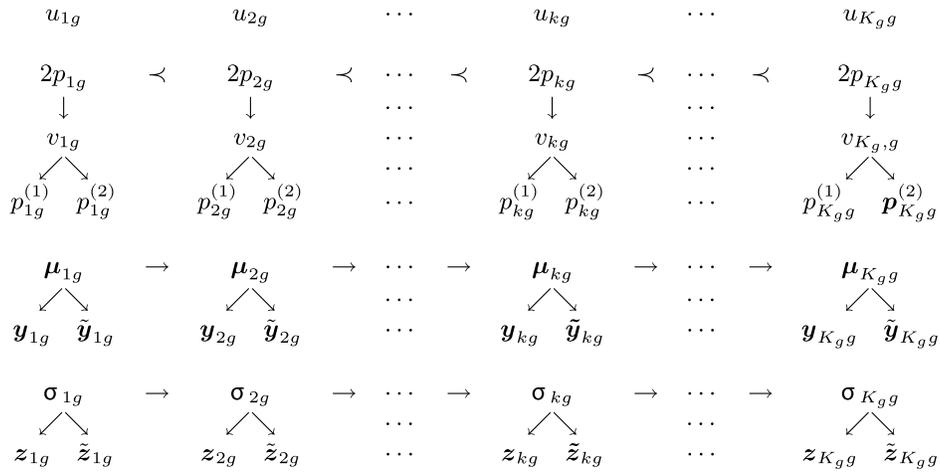}

\caption{Splits of $2\mathbf{p}_{g}$,
$\bolds{\mu}_{g}$ and $\bolds{\sigma}_{g}$. The partial
ordering $\prec$ is the ordering on $\mu_{kg}^{(1)}$s. The right
arrows ``$\rightarrow$'' represents the sequential split for
$\bolds{\mu}_{g}$ and $\bolds{\sigma}_{g}$.}\label{fig:splitp}
\end{figure}

\item[Step 3:] The aim now is to split each component of the
triplet $(\mathbf{p}_g,\bolds{\mu}_g,\bolds{\sigma}_g)$
into 2 parts: $(\mathbf{p}_{g_1},\bolds{\mu}_{g_1},
\bolds{\sigma}_{g_1})$ and
$(\mathbf{p}_{g_2},\bolds{\mu}_{g_2},
\bolds{\sigma}_{g_2})$ such that both $\bolds{\mu}_{g_1}$
and $\bolds{\mu}_{g_2}$ satisfy the constraints
(\ref{muconstraintvector}) for $g=g_1$ and $g_2$. The case of
$K_{g_1}+K_{g_2}=2K_{g}$ is first considered. A sketch of the split
move is best described by the diagram in Figure \ref{fig:splitp},
which introduces the additional variables to be used for performing
the split. In Figure \ref{fig:splitp}, $2\mathbf{p}_{g}$ is
considered for splitting because the two split components will
represent the second level mixing probabilities of $g_1$ and $g_2$,
the sum of which together equals~2.
\end{longlist}}

For each $k$, the variable $u_{kg}$ in Figure \ref{fig:splitp} takes
three values, namely, $0$, $1$ and $2$ that respectively determines
if the split components of $2{p}_{kg}$, $\bolds{\mu}_{kg}$ and
$\bolds{\sigma}_{kg}$ either (1) both go to component $g_2$, (2)
one goes to component $g_1$ and the other goes to~$g_2$, or (3) both
go to $g_1$. The variables $u_{kg}, k=1,2,\ldots,K_g$ must satisfy
several constraints: (1) $\sum_{k=1}^{K_g} u_{kg} = K_{g_1}$, (2)
$u_{kg} = 1$ for any $k$ such that $p_{kg} > 0.5$, and (3)
$\sum_{k\dvtx u_{kg} =h}2p_{kg} < 1$ for $h=0,2$. The reader is referred
to our technical report\vspace*{1pt} Dass and Li (\citeyear{DaLi2008}) for further explanation of
these restrictions.

To generate the vector $\underline{u}\equiv
(u_{1g},u_{2g},\ldots,u_{{K_g}g})'$, we consider all combinations of
$\underline{u} \in\{0,1,2\}^{K_{g}}$, and reject the ones that do
not satisfy the three restrictions. From the total number of
remaining admissible combinations, $M_1$, say, we select a vector
$\underline{u}$ randomly with equal probability $1/M_1$.

Once $\underline{u}$ has been generated, a random vector
$\underline{v}\equiv(v_{kg},  k=1,2,\ldots,K_g)$ is generated to
split $2\mathbf{p}_{g}$ (see Figure \ref{fig:splitp}). Some
notation are in order: Let $A_0 = \{ k\dvtx u_{kg} = 0\}$, $A_1 =
\{ k\dvtx u_{kg} = 1\}$ and $A_2 = \{ k\dvtx u_{kg} = 2\}$. As in the
case of $\underline{u}$, a few restrictions also need to be placed on
the vector $\underline{v}$. To see what these restrictions are, we
denote
%
\begin{equation}
\label{vsplitprob}
 p_{kg}^{(1)} = 2v_{kg}p_{kg}\quad\mbox{and}\quad
p_{kg}^{(2)} = 2(1-v_{kg})p_{kg}
\end{equation}
for $k=1,2,\ldots,K_g$, to be the split components from $2 p_{kg}$.
Note that depending on the value of $u_{kg}=0,1$ or $2$, the split
components, $p_{kg}^{(1)}$ and $p_{kg}^{(2)}$, are either both
assigned to component $g_2$, one to $g_1$ and the other to $g_2$, or
both to $g_1$. For the case $u_{kg}=1$, we will assume that
$p_{kg}^{(1)}$ is the split probability that goes to $g_1$ and
$p_{kg}^{(2)}$ goes to $g_2$. Note that the mixing probabilities for
both components $g_1$ and $g_2$\vspace*{1pt} should equal 1. This implies
\begin{equation}
\label{prob1}
\sum_{k\dvtx k\in A_{1}} p_{kg}^{(1)} + \sum_{k\dvtx k
\in A_{2}} 2 p_{kg} = 1\quad\mbox{and} \quad\sum_{k\dvtx k \in
A_1} p_{kg}^{(2)} + \sum_{k\dvtx k\in A_0} 2 p_{kg} = 1
\end{equation}
 for
components $g_1$ and $g_2$, respectively. The second equation of
(\ref{prob1}) is\break redundant if the first is assumed since
$\sum_{k\dvtx k\in A_{1}} p_{kg}^{(1)} + \sum_{k\dvtx k \in
A_{2}} 2 p_{kg} +\break \sum_{k\dvtx k \in A_1} p_{kg}^{(2)} +
\sum_{k\dvtx k\in A_0} 2 p_{kg} = 2\sum_{k=1}^{K_g} p_{kg} = 2$. We
rewrite the first equation as
\begin{equation}
\label{prob1rewritten}
\sum_{k\dvtx k \in A_1} a_{k}v_{kg} = 1,
\end{equation}
 where $a_k =
2p_{kg}/(1-\sum_{k\dvtx k\in A_2} 2p_{kg})$. Equation
(\ref{prob1rewritten}) implies that the entries of the vector
$\underline{v}$ are required to satisfy two restrictions: (1) $0 \le
v_{kg} \le1$ for $k=1,2,\ldots,K_g$ from (\ref{vsplitprob}), and (2)
equation (\ref{prob1rewritten}) above. In Dass and Li (\citeyear{DaLi2008}), an
algorithm is given to generate such a $\underline{v}$ where the
proposal density can be written down in closed form.

The split of $\bolds{\mu}_{g}$ and $\bolds{\sigma}_{g}$ is
carried out by generating two new random vectors
$\mathbf{y}_{kg}$ and $\mathbf{z}_{kg}$, for
$k=1,2,\ldots,K_g$; see Figure \ref{fig:splitp}. The generation of
$\mathbf{y}_{kg}$ is subject to restrictions arising from
constraint (\ref{muconstraintvector}) on $\bolds{\mu}_{g}$. The
other component of the split of $\bolds{\mu}_{g}$ and
$\bolds{\sigma}_{g}$, $\tilde{\mathbf{y}}_{kg}$ and
$\tilde{\mathbf{z}}_{kg}$, are obtained by solving two
(vectorized) linear equations [see Dass and Li (\citeyear{DaLi2008})]. Our techical
report also gives further details of the RJMCMC sampler, including
obtaining the new first and second level labels as well as the
deriving explicit expressions for the allocation probabilities and
the Jacobian of the transformation from
$(\bolds{\theta},\mathbf{u})$ to $\bolds{\theta}^{*}$.
\end{appendix}

\printaddresses

\end{document}